\documentclass{ws-procs975x65}

\begin{document}

\title{Holographic Estimate of Muon $g-2$\footnote{Talk given at SCGT09, 
Dec. 8-11, 2009, Japan and 
to be published by World Scientific Publishing Co., Singapore (eds. M. Harada, H. Fukaya, M.Tanabashi and K. Yamawaki).  }}

\author{Deog Ki Hong$^*$}

\address{Department of Physics, Pusan National University,\\
Busan 609-735, Korea\\
$^*$E-mail: dkhong@pusan.ac.kr}

\begin{abstract}
 I present recent calculations of the hadronic contributions to muon anomalous magnetic moment in holographic QCD, based on gauge/gravity duality. The holographic estimates are compared well with the analysis based on recently revised  BaBar measurements of $e^{+}e^{-}\to\pi^{+}\pi^{-}$ cross-sections and also with other model calculations for the light-by-light scattering contributions.  
\end{abstract}


\bodymatter

\section{Introduction}\label{aba:sec1}
The  successful prediction of anomalous magnetic moments in early days of particle physics
was one of the first triumphs of quantum field theory~\cite{Schwinger:1951nm}. Since then it has served as a critical test of the standard model of particle physics,
which is based on quantum field theory, guided by gauge principle.  Recent measurement of anomalous magnetic moment of muon (E821 at BNL)~\cite{Bennett:2006fi} provides the most stringent test so far, at the level of sub parts per million,  of the standard model, 
 \begin{equation}
 a_{\mu}^{\rm exp}=\frac{g_{\mu}-2}{2}=11659208.0(5.4)(3.3)\times 10^{-10} \,,
  \end{equation}
which exhibits currently $3.2\sigma$ deviation from the standard model estimate:
 \begin{equation}
   \Delta a_{\mu}=a_{\mu}^{\rm exp}-a_{\mu}^{\rm SM}=302(63)(61)\times 10^{-11}\,.
   \end{equation}
If the discrepancy persists in more improved measurements, such as the E969 experiment, planned at BNL, or other experiments  at the Fermi Lab,
it will certainly indicate a hint of new physics. Therefore it is quite necessary to understand the standard model predictions more precisely to pin down the possible hint of new physics at the level of $5\sigma$ or more. 

The theoretical prediction of muon $g-2$ in the standard model consists of three different contributions:
 \begin{equation}
   a_{\mu}^{\rm th}=a_{\mu}^{\rm QED}+a_{\mu}^{\rm weak}+a_{\mu}^{\rm had}\,.\nonumber
   \end{equation}
Among them the QED contribution is most dominant one and has been calculated  to be 
$a_{\mu}^{\rm QED}=116 584 718.10(0.16)\times 10^{-11}$ at 4.5 loops by Kinoshita et al~\cite{Aoyama:2008hz}, while 
the weak interaction corrections are found to be $a_{\mu}^{\rm weak}=154(2)\times 10^{-11} $ 
at the two-loop level~\cite{Czarnecki:1995sz}. 

As hadrons (or quarks) contribute to the anomalous magnetic moment of muon only through
quantum fluctuations, the strong interaction contribution is suppressed, compared to that of QED. 
The current estimate of the hadronic contributions is 
\begin{equation}
 a_{\mu}^{\rm had}=116591778(2)(46)(40)\times 10^{-11}\,.  \label{had}
 \end{equation}
The most uncertainty in the SM estimate is however coming from the hadronic contributions, since we poorly understand  strong dynamics, while the electroweak corrections can be calculated accurately in perturbation.  Direct calculation of hadronic contributions from QCD  requires lattice calculations, which are currently not accurate enough to give a meaningful result due to large systematic uncertainties~\cite{Blum:2002ii}. 

In this talk I present new estimates of the hadronic contributions, based on a holographic model of QCD. 
Holographic models  have been proposed recently for QCD~\cite{Sakai:2004cn,Erlich:2005qh}, inspired by the gauge/gravity duality, found in string theory~\cite{Maldacena:1997re}, which states that certain strongly coupled gauge theories are equivalent to weakly coupled gravity in one-higher dimension, the extra dimension being the renormalization scale. 
The holographic models of QCD were found to be quite successful to account for the properties of hadrons at low energy and give relations to their couplings and also new sum rules~\cite{Sakai:2004cn,Erlich:2005qh,Hong:2006ta,Hong:2007kx}, offering theoretical understanding of phenomenological rules for hadrons found in 60's such as vector meson dominance, proposed by Sakurai~\cite{Sakurai:1969ss}. 

\section{Holographic QCD}\label{aba:sec2}
Holographic QCD (hQCD) is a model for 5D gravity dual theory of QCD in the large $N_{c}$ and large 't Hooft coupling limit ($\lambda\equiv g_{s}^{2}N_{c}\gg1$), describing QCD directly with hadrons. 
As a gravity dual of QCD with three light flavors we consider $U(3)_{L}\times U(3)_{R}$ flavor gauge theory in a slice of $AdS_{5}$,
\begin{equation}
S=\int d^5x \sqrt{g}\textrm{ Tr}\left\{|DX|^2+3|X|^2-\frac{1}{4g_5^2}({F_L}^2+{F_R}^2)\right\}+S_{Y}+S_{CS}\,.
\label{bulkaction}
\end{equation}
where the metric is given as, taking the AdS radius $R=1$,
\begin{equation}
ds^2=\frac{1}{z^2}(dx^\mu dx_\mu-dz^2),\quad \epsilon\le z\le z_{m}\,.
\end{equation}
We take the ultraviolet (UV) regulator $\epsilon\to0$ and introduce an infra-red (IR) brane  at $z_{m}$ to implement the confinement of QCD, breaking the conformal symmetry~\cite{Erlich:2005qh}. 
The bulk scalar ($X$) and the bulk gauge fields ($A$) are dual to $\bar q\,q$ and $\bar q\gamma^{\mu}T^{a}q$, respectively. 
Following Katz and Schwartz~\cite{Katz:2007tf}, we  have introduced a flavor-singlet bulk scalar ($Y$) for the $\eta^{\prime}$ meson,  which is dual to $F^{2}$ ($F\tilde F$) and is described by
\begin{equation}
 S_{Y}=\int\,d^{5}x\sqrt{g}\left[\frac{1}{2}|DY|^2-\frac{\kappa}{2}(Y^{N_f}\textrm{det}(X)+\textrm{h.c.})\right]\label{5}\,.
\end{equation} 
Finally we introduce a Chern-Simons term to reproduce the QCD flavor anomaly~\cite{Domokos:2007kt},
\begin{equation}
S_{CS}=\frac{N_{c}}{24\pi^{2}}\int\left[\omega_{5}(A_{L})-\omega_{5}(A_{R})\right]\,,\label{cs}
\end{equation}
where $d\,\omega_{5}(A)={\rm Tr}\,F^{3}$.  
Our calculations can be easily applied to other holographic models like Sakai-Sugimoto model, which will give similar results.

Gauge/gravity duality at large $N_{c}$ and large $\lambda$ implies that the generating functional  of one-particle irreducible Green's functions is given by the classical gravity action of the dual theory:
\begin{equation}
W_{4D}[\phi_{0}(x)]=S_{5D{\rm eff}}[\phi(x,\epsilon)]\quad{\rm with}\quad \phi(x,\epsilon)=\phi_{0}(x)\,.\label{ads-cft}
\end{equation}
Using this equality we will be able to calculate the hadronic contributions at the leading order in $1/N_{c}$ and $1/\lambda$ expansion. 

\section{Holographic calculations of hadronic contributions}\label{aba:sec3}
The strong interaction contributions to the muon magnetic moment consist of three pieces;
the hadronic vacuum polarization (HLO),  the higher-order hadronic vacuum-polarization effect, 
and the hadronic light-by-light (LBL) scattering.  

The HLO contribution (see Fig.~\ref{hlo}) is given as~\cite{Blum:2002ii}
\begin{equation} 
 a_\mu^{\rm HLO} 
  =4 \pi^2 \left(\frac{\alpha}{\pi}\right)^2 \int_0^\infty 
  d Q^2 f(Q^2) \,  \bar{\Pi}_{\rm em}^{\rm had} (Q^2) 
  \,,  \label{g-2}
\end{equation}
where the vacuum polarization is defined as 
$\bar\Pi_{\rm em}^{\rm had}(Q^{2})=\Pi_{\rm em}^{\rm had}(Q^{2}) -\Pi_{\rm em}^{\rm had}(0)$
and the kernel is 
\begin{equation}
 f(Q^2) =
  \frac{m_\mu^2 Q^2 Z^3 (1-Q^2 Z)}{1 + m_\mu^2 Q^2 Z^2}
\quad{\rm with}\;\;
Z =
- \frac{Q^2 - \sqrt{Q^4 + 4 m_\mu^2 Q^2}}{2 m_\mu^2 Q^2} \,.
\end{equation}
\begin{figure} 
\centering
\psfig{file=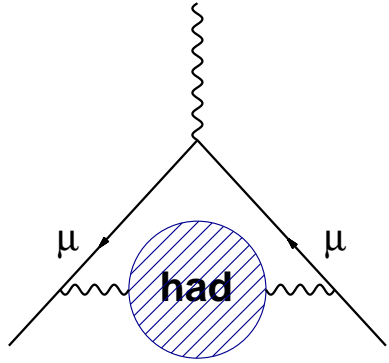,width=2in} \caption{\label{hlo}Leading hadronic-vacuum-polarization contribution to muon $g-2$.} \label{hlo}
\end{figure}

\begin{figure} 
\centering\psfig{file=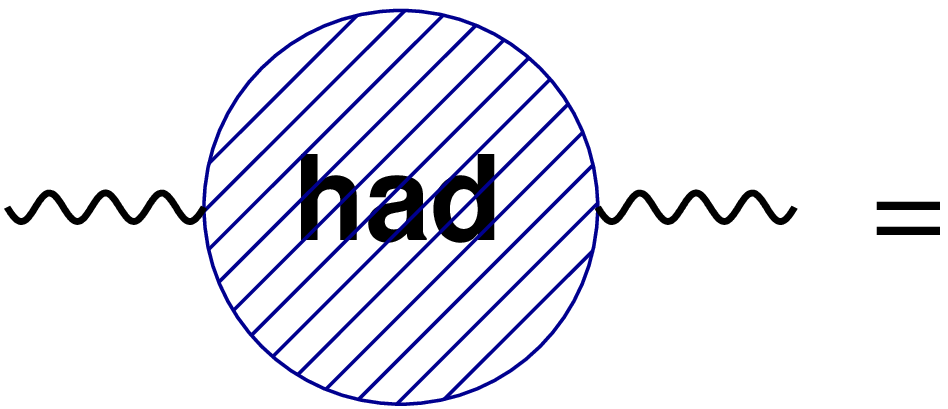,width=2in} 
\psfig{file=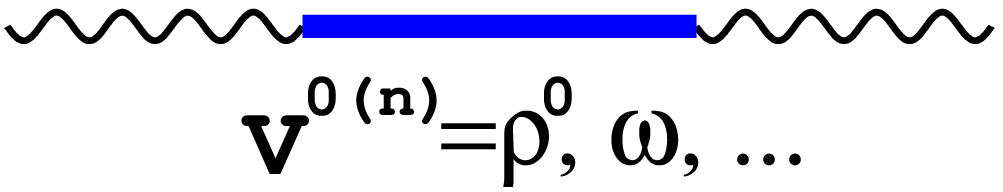,width=2in}
\caption{\label{pivfig}A diagram illustrating neutral vector meson exchanges giving dominant contributions to $\Pi_V$ at the large $N_c$ limit.}
		\label{vmd}
\end{figure}

From the AdS/CFT formula (\ref{ads-cft})  
the vector current correlator $\Pi_V(q^2)$ 
can then be expressed in terms of the infinite set of the vector meson wave-functions  $\psi_{V_n} (z)$, as shown in Fig.~\ref{vmd}, which is the holographic realization of the vector meson dominance proposed by Sakurai, 
\begin{equation} 
   \Pi_V(q^2) =  \frac{1}{g_5^2}
\sum_{n=1}^\infty \frac{[\dot{\psi}_{V_n}(\epsilon)/\epsilon]^2}{(q^2 -  M_{V_n}^2) M_{V_n}^2} 
   \,,  \label{piv:res}
\end{equation}
where the dot denotes the derivative with respect to $z$, $\dot{\psi}\equiv\partial_{z}\psi$.
Using the holographic renormalization to take care of the UV divergence and keeping only the first four low-lying states, we find for Euclidean momentum $Q^{2}=-q^{2}\ge0$~\cite{Hong:2009jv}
\begin{equation} 
 \bar{\Pi}_V(Q^2) 
  \simeq
  \sum_{n=1}^4 \frac{Q^2 F_{V_n}^2}{(Q^2 + M_{V_n}^2)M_{V_n}^4} 
  +   {\cal O}(Q^2/(M_{V_5}^2))
\, \label{bar:pi}
\end{equation} 
where $F_{V_{n}}$ is the decay constant of $n$-th excited vector mesons.
\begin{figure} 
\centering
\psfig{file=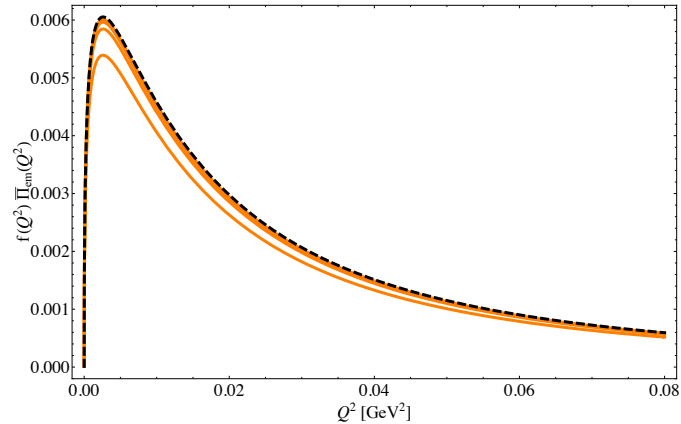,width=3.5in} 
\caption{\label{saturation}Comparison of the integral kernel 
$f(Q^2) \bar{\Pi}_{\rm em}(Q^2)$ which has a peak around $Q^2 = m_\mu^2$: 
The dashed curve corresponds to 
the result including full contributions from the infinite tower of vector mesons, 
while four bold curves are obtained by integrating out the infinite tower 
at the levels of $n=1,2,3,4$. 
The dashed curve is almost (within about 1\% deviation) reproduced when $n=4$.  
In the plot the number of flavors $N_f$ is taken to be 2.} 
\end{figure}
Plugging the holographic vacuum polarization (\ref{bar:pi}) into the formula (\ref{g-2}), we obtain~\cite{Hong:2009jv} 
\begin{equation} 
 a_\mu^{\rm HLO}|^{N_f=2}_{\rm AdS/QCD} 
=470.5 \times 10^{-10}, 
\label{U2:result}
\end{equation} 
which agrees, within 30\% errors, 
with the currently updated value~\cite{Davier:2009zi}, estimated  from new 2009 BaBar data~\cite{:2009fg}
\begin{equation} 
a_\mu^{\rm HLO}[\pi\pi] |_{\rm BABAR} 
=(514.1 \pm 3.8 )\times 10^{-10}
\,. \label{BABAR}
\end{equation}

\begin{figure} 
\centering
\psfig{file=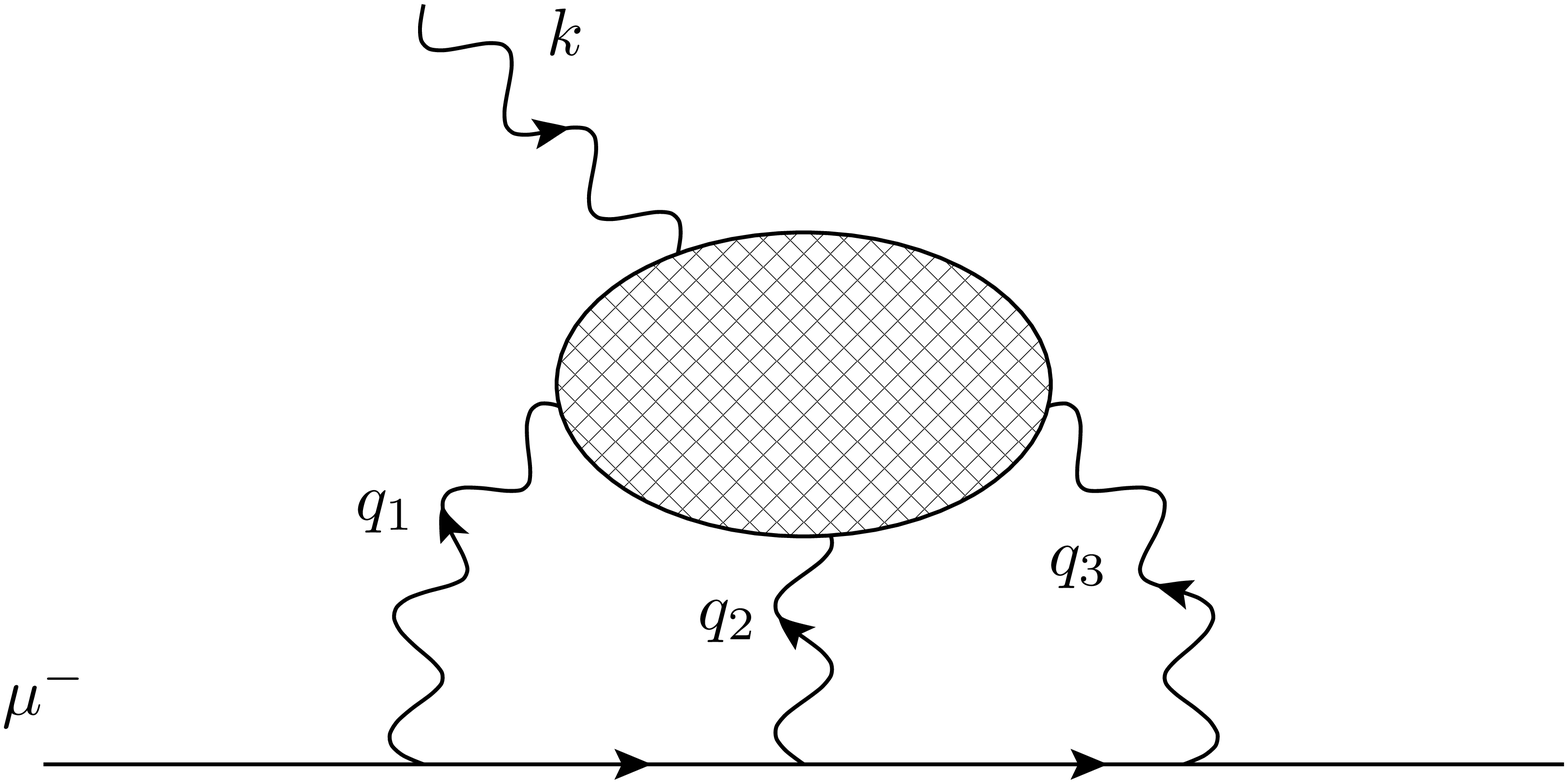,width=4in} 
\caption{\label{LBL diagram} Light-by-light corrections to muon $g-2$. }
\end{figure}

For the hadronic light-by-light corrections, shown in Fig.~\ref{LBL diagram}, 
we need to calculate 4-point functions of flavor currents.
Since there is no quartic term for $A_{Q_{\rm em}}$ ($Q_{\rm em}=1/2+I_{3}$),
there is no 1PI 4-point function for the EM currents in hQCD,
\begin{figure} 
\centering
\psfig{file=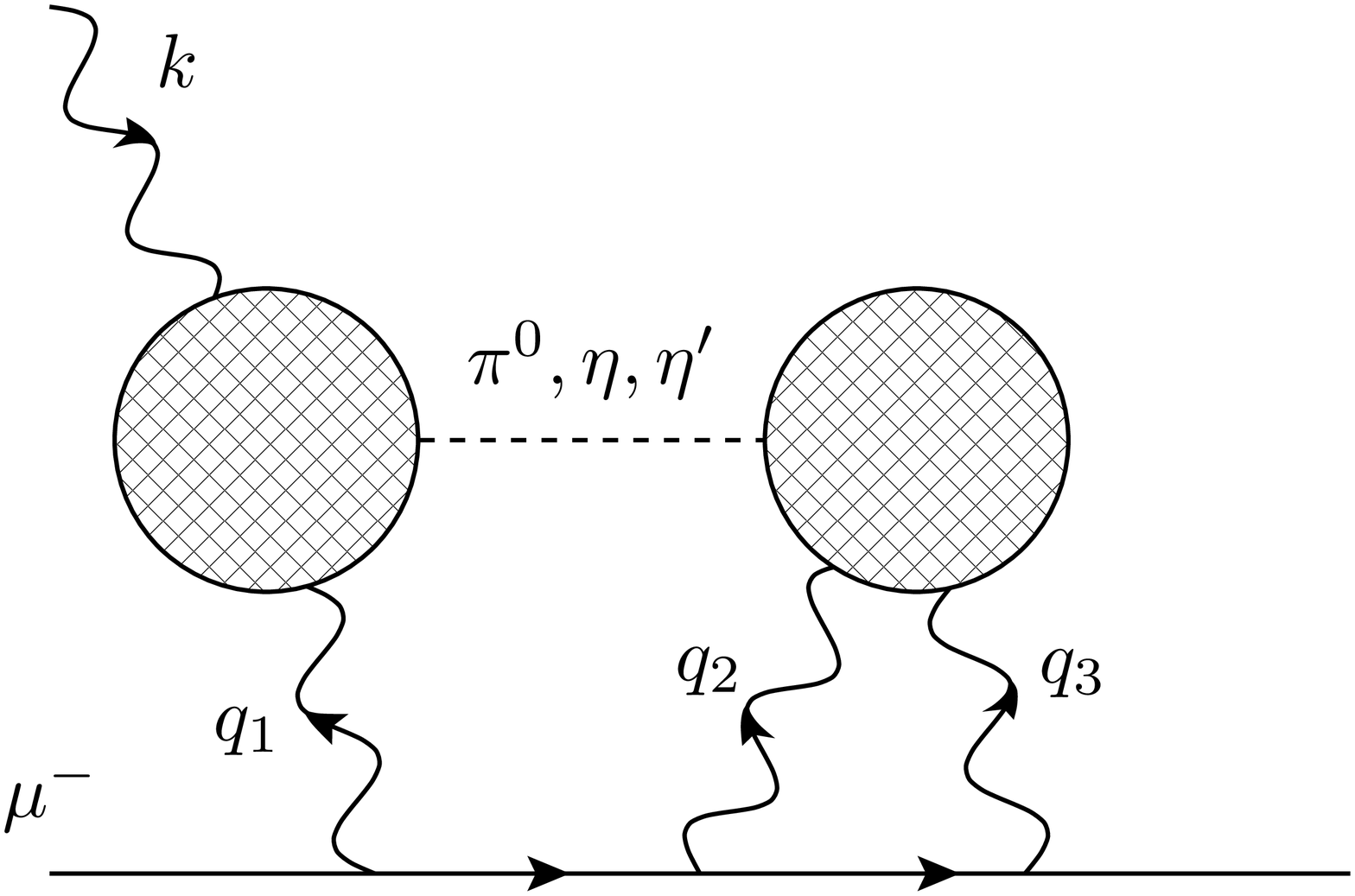,width=1.5in} 
\psfig{file=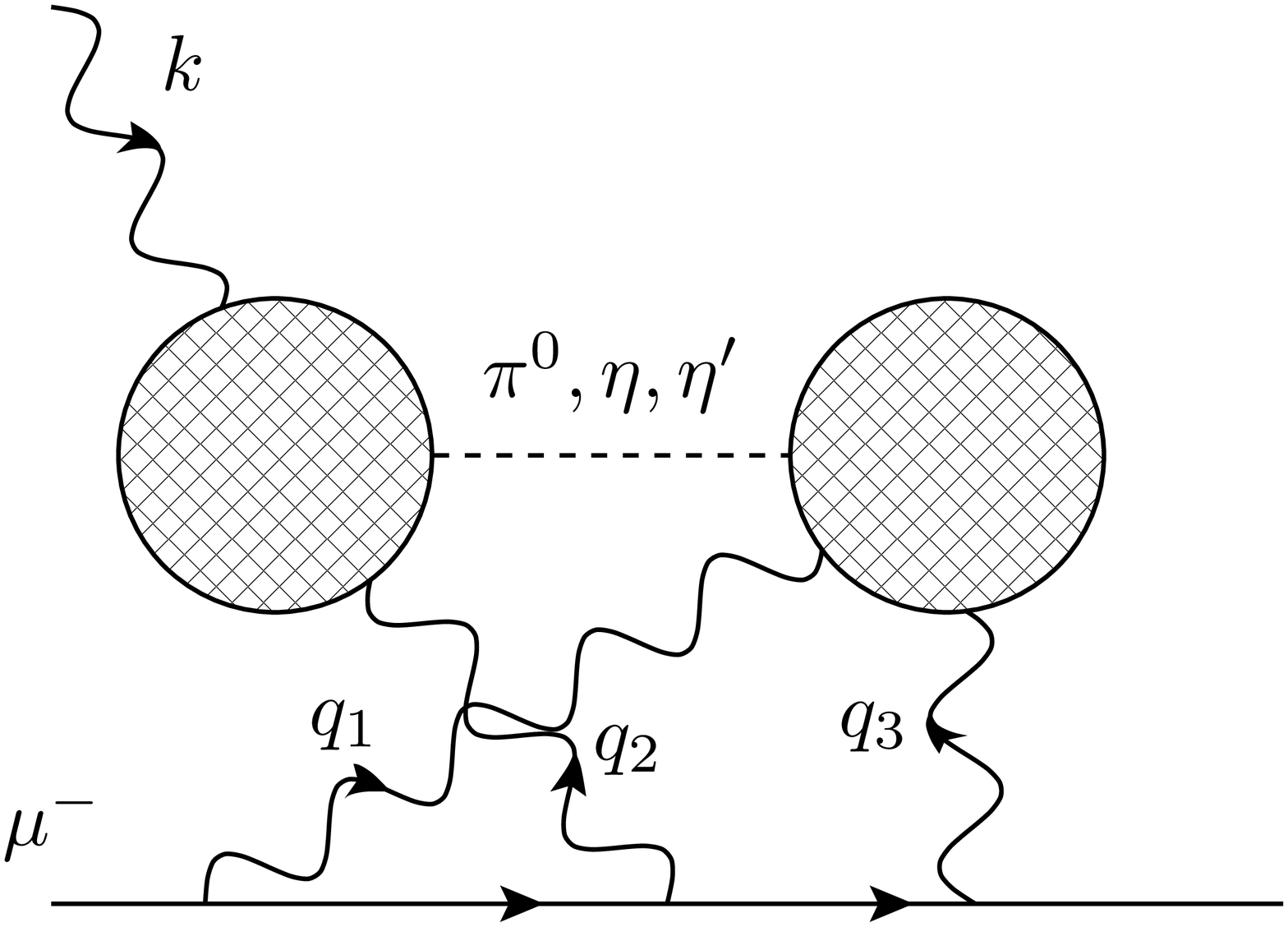,width=1.5in}
\psfig{file=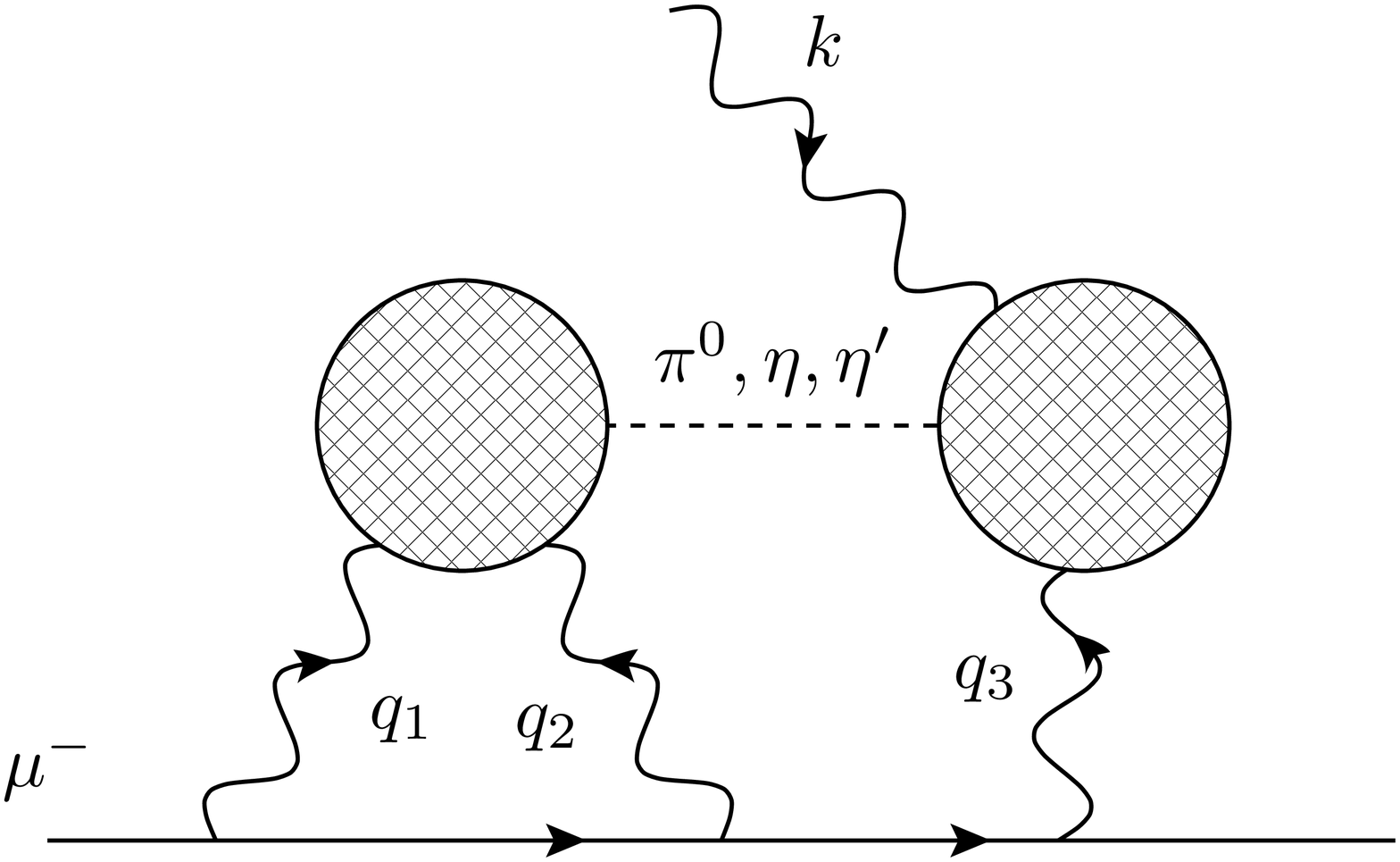,width=1.5in}
\caption{\label{LBL diagram1} Light-by-light correction  is  dominated by the pseudo scalar mesons  exchange.}
\end{figure}
because higher order terms like $F^{4}$ or $F^{2}X^{2}$ terms are suppressed.
In hQCD the LBL diagram is therefore dominated by VVA or VVP diagrams (see Fig.~\ref{LBL diagram1}),
which come from the Chern-Simons term, Eq.~\ref{cs}:
\begin{equation}
F_{\gamma^{*}\gamma^{*}P(A)}(q_{1},q_{2})=\frac{\delta^{3}}{\delta V(q_{1})\delta V(q_{2})\delta A(-q_{1}-q_{2})}S_{CS}
\end{equation}
where the gauge fields satisfy in the axial gauge, $V_{5}=0=A_{5}$, 
\begin{eqnarray}
	\left[\partial_z\left(\frac{1}{z}\partial_zV_\mu^{\hat a}(q,z)\right)+\frac{q^2}{z}V_\mu^{\hat a}(q,z)\right]_{\perp}&=&0\,,\label{1}\\
	\left[\partial_z\left(\frac{1}{z}\partial_zA_\mu^{\hat a}\right)+\frac{q^2}{z}A_\mu^{\hat a}-\frac{g_5^2v^2}{z^3}A_\mu^{\hat a}\right]_{\perp}&=&0\,,\label{2}
\end{eqnarray}
where $\perp$ denotes the projection onto the transversal components. 
For two flavors the longitudinal components, $A^{a}_{\mu\parallel}=\partial_{\mu}\phi^{a}$, and the phase of bulk scalar $X$ are related by EOM as
\begin{eqnarray}
	\partial_z\left(\frac{1}{z}\partial_z\phi^a\right)+\frac{g_5^2v^2}{z^3}(\pi^a-\phi^a)&=&0\,,\label{3}\\
	-q^2\partial_z\phi^a+\frac{g_5^2v^2}{z^2}\partial_z\pi^a&=&0\,.\label{4}
\end{eqnarray}
The anomalous form factor is then given as~\cite{Grigoryan:2008up}, with $\psi^{a}(z)=\phi^{a}-\pi^{a}$ and $J_{q}=V(iq,z)$
\begin{equation}
	F_{\pi\gamma^*\gamma^*}=\frac{N_c}{12\pi^2}\left[\psi(z_m)J(Q_1,z_m)J(Q_2,z_m)-\int_{z} \,\partial_{z}\psi J_{Q_1}J_{Q_2}\right],\label{pionff}
\end{equation}
where we take as boundary conditions
$\psi(\epsilon)=1$ and $\partial_{z}\psi(z)|_{z_{m}}=0$.
\begin{figure}[tbh]
\vskip -0.2in
	\centering
	\includegraphics[width=0.55\textwidth]{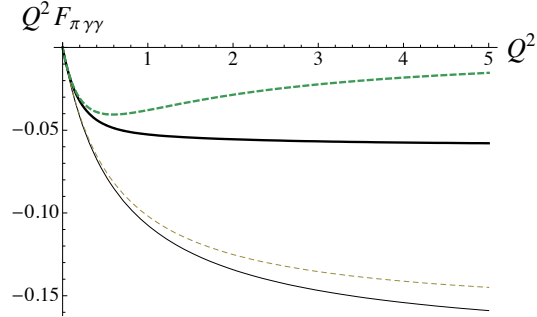}%
	\caption{ $F_{\pi\gamma^{*}\gamma}(Q^{2},0)$ for lower part; $F_{\pi\gamma^{*}\gamma^{*}}(Q^{2},Q^{2})$  for upper part (Brodsky-Lepage).}
\label{qcd}
\end{figure}

To calculate the hadronic LBL contribution to the muon anomalous magnetic moment we expand the photon line in the anomalous form factor (\ref{pionff}) as 
\begin{equation}
	J(-iQ,z)=V(q,z)=\sum_{\rho}\frac{-g_5f_{\rho}\psi_\rho(z)}{q^2-m_\rho^2+i\epsilon}
	\label{modeexpansion}\nonumber
\end{equation}
The result is shown in Table~\ref{result} for several choices of vector-mode truncations~\cite{Hong:2009zw}.
\begin{table}
\tbl{Muon $g-2$ in unit of $10^{-10}$ from AdS/QCD.}
{\begin{tabular}{@{}cccccc@{}}
\toprule
Vector modes & \hphantom{0}$a_\mu^{\pi^0}$ \hphantom{0}& \hphantom{0}$a_\mu^{\eta}$\hphantom{0} &\hphantom{0}$a_\mu^{\eta^\prime}$\hphantom{0}& & \hphantom{00}$a_\mu^{\textrm{PS}}$\hphantom{00}\\
 \colrule
4 & 7.5\hphantom{0}& 2.1 & 1.0&&10.6 \\
6 & 7.1 \hphantom{0}& 2.5 & 0.9 && 10.5 \\
8 & 6.9\hphantom{0}& 2.7& 1.1 & &10.7\\\botrule
\end{tabular}}
\label{result}
\end{table}
Our results are comparable with recent results by A. Nyffeler~\cite{Nyffeler:2009tw},  obtained in the LMD+V model;
\begin{equation}
a_{\mu}^{\rm PS}=9.9(1.6)\times10^{-10}\,.
\end{equation}

\section{Discussions}\label{aba:sec4}
In the era of electroweak precision, it becomes more important to understand precisely QCD corrections to the electroweak processes. As being non-perturbative strong dynamics, QCD corrections are often difficult to estimate and one resorts to lattice calculations, which are, however,  not precise enough for certain measurements such as muon anomalous magnetic moment. Recent development in gauge/gravity duality shows that in the large $N_{c}$ and large $\lambda$ limit the estimate of QCD contributions can be made precisely in holographic QCD, thus may be useful in assessing the new physics effects. I present recent estimates~\cite{Hong:2009jv,Hong:2009zw} of anomalous magnetic moment of muon in holographic QCD, which are found to be in consistent with other calculations.

\section*{Acknowledgments}
D.K.H. 
is grateful to the organizers of SCGT09 for  a very stimulating  meeting and thanks Doyoun Kim and Shinya Matsuzaki for the collaborations upon which this talk is based on. 
This work is supported by the Korea Research Foundation Grant funded by the Korean Government (MOEHRD, Basic Research Promotion Fund) (KRF-2007-314- C00052).

\end{document}